\documentclass[prl,twocolumn,showpacs,preprintnumbers,amsmath,amssymb]{revtex4}

\usepackage{graphicx}
\usepackage{dcolumn}
\usepackage{bm}

\begin{document}


\title{Effects of Rattling Phonons on the 
Quasiparticle Excitation and Dynamics 
in the Superconducting $\beta$-Pyrochlore KOs$_2$O$_6$}

\author{Y. Shimono,$^1$ T. Shibauchi,$^{1}$ Y. Kasahara,$^{1}$ 
T. Kato,$^{1}$ K. Hashimoto,$^{1}$ Y. Matsuda,$^{1,2}$ \\
J. Yamaura,$^2$ Y. Nagao,$^2$ and Z. Hiroi$^2$}

\affiliation{$^1$Department of Physics, Kyoto University,
Sakyo-ku, Kyoto 606-8502, Japan\\
$^2$Institute for Solid State Physics, University of Tokyo,
Kashiwa, Chiba 277-8581, Japan}

\date{\today}

\begin{abstract}
Microwave penetration depth $\lambda$ and surface resistance 
at 27~GHz are measured in high quality crystals of KOs$_2$O$_6$. 
Firm evidence for fully-gapped superconductivity is provided 
from $\lambda(T)$. 
Below the second transition at $T_{\rm p}\sim 8$~K, the superfluid 
density shows a step-like change 
with a suppression of effective critical temperature $T_{\rm c}$. 
Concurrently, the extracted quasiparticle scattering time 
shows a steep enhancement, indicating 
a strong coupling between the anomalous rattling motion of K ions 
and quasiparticles. 
The results imply that the rattling phonons 
help to enhance superconductivity, 
and that K sites freeze to an ordered state 
with long quasiparticle mean free path below $T_{\rm p}$. 

\end{abstract}


\maketitle

Recently, the roles of phonons have become refocused 
on a plethora of novel physical properties in strongly correlated 
electron systems, 
such as their interplay with superconductivity in high-$T_{\rm c}$ 
cuprates \cite{Davis}, 
and the heavy fermion behavior in filled-skutterudites \cite{Bauer} 
possibly due to unconventional motions of ions \cite{Hattori}. 
In the newly discovered $\beta$-pyrochlore superconductor 
KOs$_2$O$_6$ with relatively high superconducting critical 
temperature $T_{\rm c} \approx 9.5$~K \cite{YonezawaK}, 
the low-energy local vibration, or 
the anharmonic ``rattling'' motion of K ions with very large excursion 
inside an oversized Os-O atomic cage, has been 
demonstrated both theoretically \cite{Kunes,Kunes2006} and 
experimentally \cite{Yamaura,Hiroi,Batlogg,Hiroi_p,Kasahara,Yoshida}. 
It is believed that the pronounced rattling of K ions is responsible for 
the unusual convex temperature dependence of resistivity $\rho(T)$ 
in the normal state of KOs$_2$O$_6$ \cite{Hiroi}, indicating that 
the rattling strongly influences the electronic structure. 
It is also suggested \cite{Hattori} that such low-lying anharmonic phonons 
may give rise to an exotic superconducting state through the possible 
formation of heavy quasiparticles due to 
off-center degrees of freedom of ions. 
Very little is known, however, about effects of the rattling on the 
superconducting properties. 

What is intriguing in KOs$_2$O$_6$ is that 
inside the superconducting state below $T_{\rm c}$, 
a second transition occurs at $T_{\rm p}\sim 8$~K, 
where specific heat shows an almost field-independent anomaly 
\cite{Hiroi,Batlogg,Hiroi_p}. 
The high-field transport measurements have revealed that 
the concave $\rho(T)$ at high temperatures changes to 
a typical Fermi-liquid dependence $AT^2$ below $T_{\rm p}$ 
\cite{Kasahara,Hiroi_p}. 
This result naturally suggests that the K rattle responsible for the 
anomalous transport properties should be frozen below the transition 
$T_{\rm p}$. 
This provides a unique opportunity to study how this unusual rattling 
affects superconductivity and quasiparticle dynamics in the 
superconducting state. 

In addition to these odd behaviors, strong electron correlations appear to 
have important contributions in transport and thermodynamic properties 
of KOs$_2$O$_6$. 
Pyrochlore structure, having corner-shared tetrahedra network of 
transition metal that geometrically induces magnetic frustration, 
has been a platform of various novel physics such as 
heavy fermion behavior without $f$ electrons \cite{Kondo}. 
Sommerfeld coefficient $\gamma$ in KOs$_2$O$_6$ 
is estimated as high as 70 to 110 
mJ K$^{-2}$ mol$^{-1}$ \cite{Hiroi_p,Batlogg}, which is largely 
enhanced from the band calculation value of 9.8 mJ K$^{-2}$ mol$^{-1}$ 
\cite{Kunes}. 
At the superconducting transition the specific heat shows a large jump 
$\Delta C/T_c \approx 200$~mJ K$^{-2}$ mol$^{-1}$, 
and the upper critical field is found to be very high (up to 32~T 
in the zero temperature limit) with a steep slope 
of ${\rm d}H_{c2}(T)/{\rm d}T = -3.4$ T/K \cite{Shibauchi}. 
These results indicate the pairing of electrons with enhanced effective 
mass $m^*$, which in many cases invokes unconventional superconductivity. 
Furthermore, the coefficient $A$ in the $AT^2$ dependence of $\rho$ 
below $T_{\rm p}$ is found to follow the Kadowaki-Woods (KW) relation 
expected for strong correlation systems having 
large $\gamma$ \cite{Kasahara,Hiroi_p}. 

To clarify the effect of rattling phonons on superconductivity, 
microwave surface impedance $Z_{\rm s}(T)$ is a powerful low-energy 
electronic probe of quasiparticles, from which the magnetic 
penetration depth $\lambda(T)$ and quasiparticle scattering time $\tau$ 
can be extracted. 
The number of excited quasiparticles is most directly related to 
$\lambda(T)$, since the superfluid density $n_{\rm s}$ 
is proportional to $\lambda^{-2}$. 
Previous measurements of $\lambda$ by $\mu$SR \cite{Koda} 
imply anisotropic gap with nodes, which contradicts the recent 
thermal conductivity results \cite{Kasahara}. 

Here we report precise measurements of 
$Z_{\rm s}(T)$ in KOs$_2$O$_6$, 
from which fully gapped superconductivity with a large gap value 
$\Delta\approx 25$~K is unambiguously demonstrated. 
In high quality crystals, the superfluid density 
shows a clear anomaly near $T_{\rm p}$, 
and the effective $T_{\rm c}$ is reduced by the freezing transition. 
This suggests that the rattling motion is helpful for 
superconductivity.  
Furthermore, the quasiparticle scattering rate rapidly decreases 
below $T_{\rm p}$, indicating enormous 
inelastic scattering in the normal state 
due to underlying strong correlations. 

The surface impedance $Z_{\rm s}=R_{\rm s}+{\rm i}X_{\rm s}$ 
is measured by a cavity perturbation method 
with the hot finger technique \cite{Sridhar}. 
We used a 27~GHz TE$_{011}$-mode superconducting Pb cavity, 
whose temperature is maintained at 1.4~K, and the sample temperature 
is controlled up to 100~K. 
The inverse of quality factor $1/Q$ and the shift in the resonance frequency 
$\omega/2\pi$ are proportional to the real and imaginary parts 
of $Z_{\rm s}$ respectively \cite{Klein}. Details of the experimental 
setup are described elsewhere \cite{ShibauchiPRL,Shibata}. 

Single crystals of KOs$_2$O$_6$ were grown by the technique described in 
Ref.~\onlinecite{Hiroi}. It has been known that once partial hydration 
takes place, the anomaly in specific heat at 
$T_{\rm p}$ tends to collapse. 
Before microwave measurements, we checked the specific heat 
anomaly and a special care was taken to keep the crystals in a dry 
atmosphere. We measured several crystals with shiny surfaces, and 
we here focus on the results of the crystal having most pronounced anomaly 
at $T_{\rm p}$. 
The skin depth at 27 GHz is much smaller than 
the sample dimensions ($0.5\times 0.5\times 0.2$~mm$^3$), which ensures 
the skin depth regime. 

\begin{figure}
\includegraphics[width=90mm]{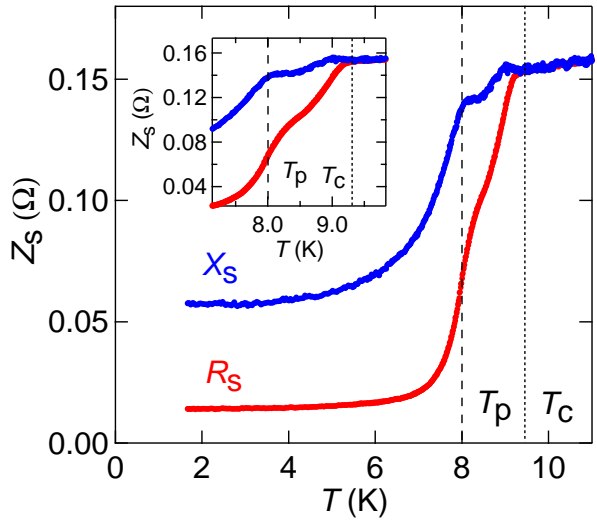}%
\caption{(color online). 
Temperature dependence of the surface resistance $R_{\rm s}$ 
and reactance $X_{\rm s}$ at 27~GHz 
in a KOs$_2$O$_6$ single crystal.
The inset shows an expanded view near the superconducting transition 
$T_{\rm c}$ and the second transition $T_{\rm p}$.}
\end{figure}

Figure~1 shows the temperature dependence of $R_{\rm s}$ and $X_{\rm s}$. 
In the normal state 
we can use the expected relation 
(in the Hagen-Rubens limit) 
$ R_{\rm s} = X_{\rm s} = (\mu_0 \omega \rho/2)^{1/2}$ 
to determine the absolute value of $Z_{\rm s}$. 
Just below $T_{\rm c}$, we observe a coherence peak in the 
reactance $X_{\rm s}(T)$ near 9~K, 
which is a typical feature in $s$-wave superconductors \cite{Klein}. 
A striking feature is that $Z_{\rm s}(T)$ shows distinct 
anomalies near $T_{\rm p}$, which we will discuss later. 

The surface reactance is proportional to the penetration depth by 
$X_{\rm s} = \mu_0 \omega \lambda$. The temperature dependence of $\lambda$ 
at low temperatures is demonstrated in Fig.~2. It is clear from the figure 
that $\lambda(T)$ has a flat temperature dependence at low temperatures, 
obviously different from $T$, $T^2$, or $T^3$ dependence \cite{Annett} 
expected in the superconducting gap function 
with line or point nodes [see inset of Fig.~2]. 
The data below 6~K can be fitted to an exponential dependence 
$\lambda(T)-\lambda(0) \propto \exp(-\Delta/k_{\rm B}T_{\rm c})$, 
with $\Delta \approx24.5$~K, giving 
a strong coupling value of $2\Delta/k_{\rm B}T_{\rm c} \approx 5.1$. 
This unambiguously indicates that the quasiparticle excitation is of 
activated type and the superconducting gap is nodeless. 
The obtained value of $\lambda(0)\approx 260$~nm is consistent with the 
$\mu$SR results \cite{Koda}, and by using the coherence length 
$\xi(0) \approx 3.2$~nm \cite{Shibauchi}, 
the Ginzburg-Landau parameter is evaluated 
as $\lambda(0)/\xi(0)\sim 82$ indicating the London limit. 
The long London penetration depth 
$\lambda_{\rm L} =(\frac{m^*}{\mu_0 n_{\rm s} e^2})^{1/2}$
gives another support for the pairing of electrons with enhanced mass. 

\begin{figure}
\includegraphics[width=90mm]{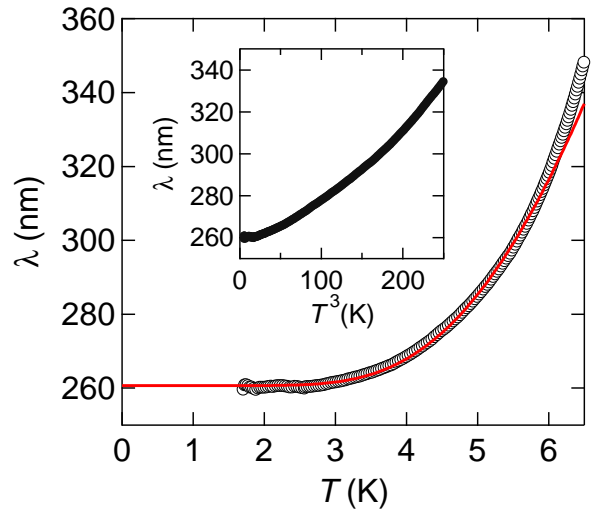}%
\caption{(color online). Temperature dependence of the microwave 
penetration depth $\lambda$ below 6.5 K in KOs$_2$O$_6$. 
The solid line is a fit to the exponential dependence of 
$\lambda(0)+C\exp(-\Delta/k_{\rm B}T_{\rm c}$) with $\lambda(0)=261$~nm 
and $\Delta=24.5$~K. The inset shows $\lambda$ vs $T^3$.}
\end{figure}

\begin{figure}
\includegraphics[width=90mm]{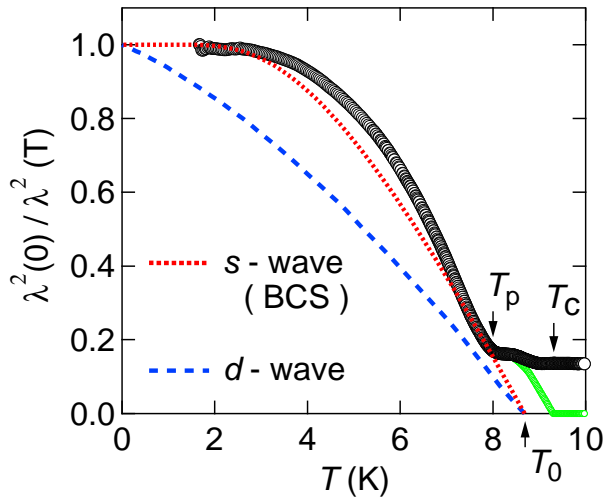}%
\caption{(color online). 
$\lambda^2(0)/\lambda^2(T) = n_{\rm s}/n$ as a function of temperature. 
The data 
below 8~K extrapolates to zero at $T_0(<T_{\rm c})$. Above $T_{\rm c}$, 
$\lambda$ is limited by the normal-state skin depth \cite{ShibauchiPRL}, 
but the superfluid density $n_{\rm s}$ should become zero at 
$T_{\rm c}$ as shown in the green solid line. 
The red dotted line is weak-coupling BCS prediction for $s$-wave 
superconductors in the London limit. The blue dashed line is 
a calculation for $d$-wave superconductors with line nodes \cite{Klemm}.} 
\end{figure}

In Fig.~3, we plot the temperature dependence of superfluid density 
$n_{\rm s}(T)/n_{\rm s}(0)=\lambda^2(0)/\lambda^2(T)$. 
Again, it is clearly incompatible with a $d$-wave calculation 
with line nodes \cite{Klemm}. We also compare the low temperature data 
with the expectation of weak-coupling BCS $s$-wave superconductors, 
and found that above 3~K the data deviates, which can be explained 
by the strong electron-phonon coupling. 
The conclusion of full gap superconductivity is reinforced by the 
observed coherence peak in $X_{\rm s}(T)$ in Fig.~1. We note that such a 
coherence peak in $X_{\rm s}(T)$ and the flat temperature dependence 
in low-temperature $\lambda(T)$ are observed in all the samples 
we measured. 
Our conclusion is also consistent with the observed weak 
field dependence of thermal conductivity $\kappa$ in the 
low-temperature limit \cite{Kasahara}. In contrast, the $\mu$SR 
paper \cite{Koda} reports on a strong field dependence of the effective 
penetration depth $\lambda_{\rm eff}$, 
but we note that it has been pointed out that the theoretical 
models employed for the data analysis may have insufficient accuracy 
\cite{Landau}, and the origin of $\lambda_{\rm eff}(H)$ is still 
controversial \cite{Sonier}. 

Next we discuss the effect of the second transition. 
The superfluid density in Fig.~3 exhibits a step-like change 
near $T_{\rm p}$. This indicates that the transition 
clearly affects the superconducting condensates. 
The temperature dependence of 
superfluid density below 8 K extrapolates to zero at a temperature 
$T_0\sim 8.7$~K 
noticeably lower than the actual $T_{\rm c}$. 
This immediately indicates that below $T_{\rm p}$ where 
the K rattle responsible for 
the anomalous $\rho(T)$ is frozen, the effective $T_{\rm c}$ 
is reduced considerably. 
We note that recent measurements of the lower critical 
field $H_{\rm c1}(T)$ \cite{Hc1}, 
which is also related to the superfluid density, show 
a similar reduction of the effective $T_{\rm c}$ 
below $T_{\rm p}$, consistent with our observation. 
These results lead us to infer that {\em the rattling motion of K ions 
helps to enhance superconductivity in this system}, 
although further theoretical investigations are necessary to 
clarify the microscopic origins of the observed behavior. 

To see the effect on the quasiparticle dynamics, 
we extract microwave conductivity 
$\sigma=\sigma_1-{\rm i}\sigma_2$ 
from the surface impedance by 
$Z_{\rm s} = (\frac{{\rm i} \mu_0 \omega}
{\sigma_1 - {\rm i} \sigma_2})^{1/2}$. 
The extracted real part of $\sigma(T)$ is plotted in Fig.~4. 
Below $T_{\rm c}$ the conductivity goes up with lowering temperature and 
below $T_{\rm p}$ it increases more rapidly. 
This dependence is markedly 
different from the usual BCS expectation in strong-coupling 
$s$-wave superconductors \cite{Klein}, 
where $\sigma_1(T)$ shows a small coherence peak just below $T_{\rm c}$ and 
it decreases rapidly below $\sim0.9T_{\rm c}$ and continues to decrease 
exponentially at lower temperatures. 
The observed enhancement of $\sigma_1(T)$ is consistent 
with the recent thermal conductivity data in a similar 
crystal \cite{Kasahara}, 
where $\kappa/T$ is enhanced in the superconducting 
state [see inset of Fig.~4]. 
Such a big enhancement in $\sigma_1(T)$ has been also reported in 
high-$T_{\rm c}$ cuprates \cite{Bonn,ShibauchiJPSJ} 
and heavy-fermion CeCoIn$_5$ superconductors \cite{Ormeno,KasaharaPRB}, 
and it has been believed as a characteristic feature of unconventional 
($d$-wave) superconductors with large electron correlation effects. 
In contrast, KOs$_2$O$_6$ is an $s$-wave superconductor and 
such a behavior is nevertheless observed, which indicates 
the uniqueness of this system. 

\begin{figure}
\includegraphics[width=90mm]{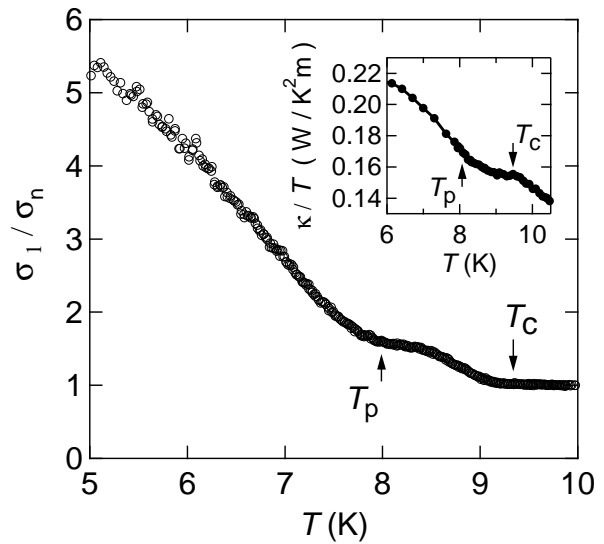}%
\caption{Temperature dependence of microwave conductivity $\sigma_1$ 
of KOs$_2$O$_6$ at 27~GHz 
normalized to the normal-state value $\sigma_{\rm n}$ at $T_{\rm c}$. 
The inset shows the $\kappa/T(T)$ data \cite{Kasahara}. }
\end{figure}

Since the effect of coherence factors appears only in the vicinity of 
$T_{\rm c}$ and the observed enhancement of $\sigma_1(T)$ 
is much bigger, we can employ the simple two-fluid analysis 
which has been known to be useful 
to evaluate the quasiparticle scattering time $\tau$ 
in the superconducting state 
\cite{Bonn,ShibauchiJPSJ}. Here, the superfluid density $n_{\rm s}$ 
and normal fluid (quasiparticle) density $n_{\rm n}$ gives 
the total carrier density $n$, 
and the real part of conductivity can be written as 
$\sigma_1 =\frac{n_{\rm n} e^2 \tau}{m^*} \frac{1}{1+(\omega \tau)^2}$. 
By using the conductivity and the superfluid density data, we get $\tau(T)$ 
as depicted in Fig.~5. 
It demonstrates that the quasiparticle scattering time is enhanced in the 
superconducting state and reaches an order of magnitude larger 
at $\sim 0.7T_{\rm c}$ than at $T_{\rm c}$. 
We note that such an enhancement of $\tau$ has been suggested by the 
thermal conductivity data \cite{Kasahara}, but one has to argue 
that the lattice contribution to $\kappa$ should be small compared 
to the electronic part. 
In contrast, the microwave conductivity is purely electronic. 

\begin{figure}
\includegraphics[width=90mm]{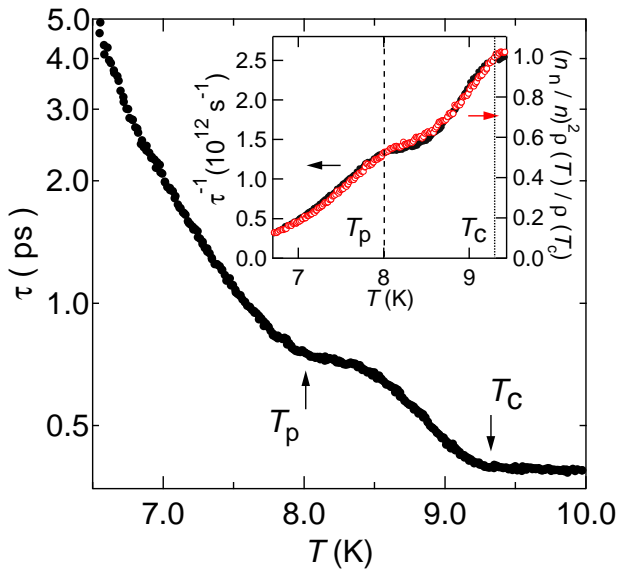}%
\caption{(color online). 
Temperature dependence of quasiparticle scattering time $\tau(T)$ 
extracted from the 
surface impedance by the two-fluid analysis. 
The inset compares $1/\tau(T)$ with 
$(n_{\rm n}(T)/n)^2\rho(T)/\rho(T_{\rm c})$.}
\end{figure}

The enhancement of $\tau$ (or the suppression of the quasiparticle 
scattering rate $1/\tau$) below $T_{\rm c}$ is an indication that there 
is enormous inelastic scattering in the normal state which is 
reduced in the superconducting state by the opening gap in the electronic 
spectrum. To see this more clearly, we compare $1/\tau$ with the quantity 
$(n_{\rm n}(T)/n)^2\rho(T)/\rho(T_{\rm c})$ in the inset of Fig.~5. 
Here $\rho(T)$ is the resistivity in the normal state obtained 
by applying strong magnetic fields (13~T) to destroy superconductivity 
\cite{Kasahara}, whose temperature dependence should be mostly come from 
the normal-state scattering rate. The factor $(n_{\rm n}(T)/n)^2$ represents 
that in the isotropic gap case the scattering rate between quasiparticles 
can be simply proportional to the number of quasiparticle itself {\em and} 
the number of scatterers. 
The temperature dependence of both quantities below $T_{\rm p}$ is 
in very good correspondence. 
This simple analysis implies that the effect of the second transition 
on the quasiparticle dynamics is two fold: (1) The normal-state 
scattering mechanism changes from strong phonon-dominated scattering 
with concave temperature dependence of $\rho(T)$ above $T_{\rm p}$ to 
strong electron-electron scattering as revealed by $AT^2$ dependence 
below $T_{\rm p}$ \cite{Hiroi_p,Kasahara}. 
(2) The number of quasiparticle in the superconducting state 
changes at $T_{\rm p}$ which also 
changes inter-quasiparticle scattering manifested by the steep 
enhancement of $\tau$. 
These effects are consistent with the views of the rattling freezing 
as the nature of the transition, and suggest 
strong electron-electron correlations inherent in this system, 
which is consistent with the observation of KW relation 
at low temperatures. 
The quasiparticle mean free path $l(T)=v_{\rm F}\tau \sim \xi\Delta\tau/\hbar$ 
reaches a long value $\sim45$~nm at $\sim 0.7T_{\rm c}$, and shows 
no saturation behavior, which may rule out disordered glass-like 
freezing below $T_{\rm p}$. 
Consistently, a theoretical suggestion that an ordered state of K sites 
appear below $T_{\rm p}$ has been made \cite{Kunes2006}. 

In summary, from the microwave surface impedance in high quality 
single crystals of KOs$_2$O$_6$, we clarify the following three points. 
(i) The superconducting ground state is fully gapped and 
electron-phonon coupling is strong. 
(ii) The superfluid density shows a step-like anomaly near the 
transition at $T_{\rm p}$, which suggests that the rattling motion 
is an important ingredient to the enhanced superconductivity.
(iii) The quasiparticle scattering is rapidly decreased 
below $T_{\rm p}$, suggesting strong coupling between 
the rattling phonons and quasiparticles. 
Our results highlight that in spite of conventional 
$s$-wave ground state, this pyrochlore superconductor with unusual 
structural and electronic properties gives a remarkable 
features in the superconducting state. An immediate question 
arises on how the transition affects vortex physics, 
which deserves further studies.

We acknowledge fruitful discussions with S. Fujimoto, M. Takigawa, 
C.~J. van~der~Beek, A.~I. Buzdin, K. Machida, and M. Sigrist. 
This work was partly supported by Grants-in-Aid for
Scientific Research from MEXT.

\end{document}